\documentclass{sigchi}

\CopyrightYear{2019}
\setcopyright{acmlicensed}
\doi{}
\isbn{}
\conferenceinfo{CHI'19,}{May 07--12, 2016, San Jose, CA, USA}
\acmPrice{\$15.00}

\toappear{
Permission to make digital or hard copies of all or part of this work
for personal or classroom use is granted without fee provided that
copies are not made or distributed for profit or commercial advantage
and that copies bear this notice and the full citation on the first
page. Copyrights for components of this work owned by others than ACM
must be honored. Abstracting with credit is permitted. To copy
otherwise, or republish, to post on servers or to redistribute to
lists, requires prior specific permission and/or a fee. Request
permissions from \href{mailto:Permissions@acm.org}{Permissions@acm.org}. \\
}

\usepackage{balance}       
\usepackage{graphics}      
\usepackage[T1]{fontenc}   
\usepackage{txfonts}
\usepackage{mathptmx}
\usepackage{color}
\usepackage{booktabs}
\usepackage{textcomp}
\usepackage[pdflang={en-US}]{hyperref}

\usepackage{microtype}        
\usepackage{ccicons}          

\def\plaintitle{Improving Usability, Efficiency, and Safety of UAV Path Planning through a Virtual Reality Interface}

\def\emptyauthor{}
\def\plainkeywords{Virtual Reality; Unmanned Aerial Vehicles; Path Planning.}

\makeatletter
\def\url@leostyle{%
  \@ifundefined{selectfont}{
    \def\UrlFont{\sf}
  }{
    \def\UrlFont{\small\bf\ttfamily}
  }}
\makeatother
\def\pprw{8.5in}
\def\pprh{11in}

\definecolor{linkColor}{RGB}{6,125,233}
\hypersetup{%
  pdftitle={\plaintitle},
  pdfauthor={\emptyauthor},
  pdfkeywords={\plainkeywords},
  pdfdisplaydoctitle=true, 
  bookmarksnumbered,
  pdfstartview={FitH},
  colorlinks,
  citecolor=black,
  filecolor=black,
  linkcolor=black,
  urlcolor=linkColor,
  breaklinks=true,
  hypertexnames=false
}

\begin{document}

\title{\plaintitle}

\numberofauthors{7}
 \author{
   Jesse Paterson, Jiwoong Han, Tom Cheng, Paxtan Laker, David McPherson, Joseph Menke, Allen Yang
}

\maketitle
\begin{abstract}
As the capability and complexity of UAVs continue to increase, the human-robot interface community has a responsibility to design better ways of specifying the complex 3D flight paths necessary for instructing them. Immersive interfaces, such as those afforded by virtual reality (VR), have several unique traits which may improve the user's ability to perceive and specify 3D information. These traits include stereoscopic depth cues which induce a sense of physical space as well as six degrees of freedom (DoF) natural head-pose and gesture interactions. This work introduces an open-source platform for 3D aerial path planning in VR and compares it to existing UAV piloting interfaces. Our study has found statistically significant improvements in safety and subjective usability over a manual control interface, while achieving a statistically significant efficiency improvement over a 2D touchscreen interface. The results illustrate that immersive interfaces provide a viable alternative to touchscreen interfaces for UAV path planning.
\end{abstract}


\category{H.5.2.}{Information Interfaces and Presentation
  (e.g. HCI)}{User Interfaces} \category{Interaction styles}{}{}

\keywords{\plainkeywords}

\section{Introduction}
Traditionally, flying unmanned aerial vehicles (UAVs) using manual joystick controls requires high levels of training and experience to manage the risk of crashing. This risk is exacerbated when flying among humans or carrying equipment for operations such as geographical survey, mechanical inspection, and disaster relief.

\begin{figure}[t]
    \centering
    \includegraphics[width=0.8\columnwidth]{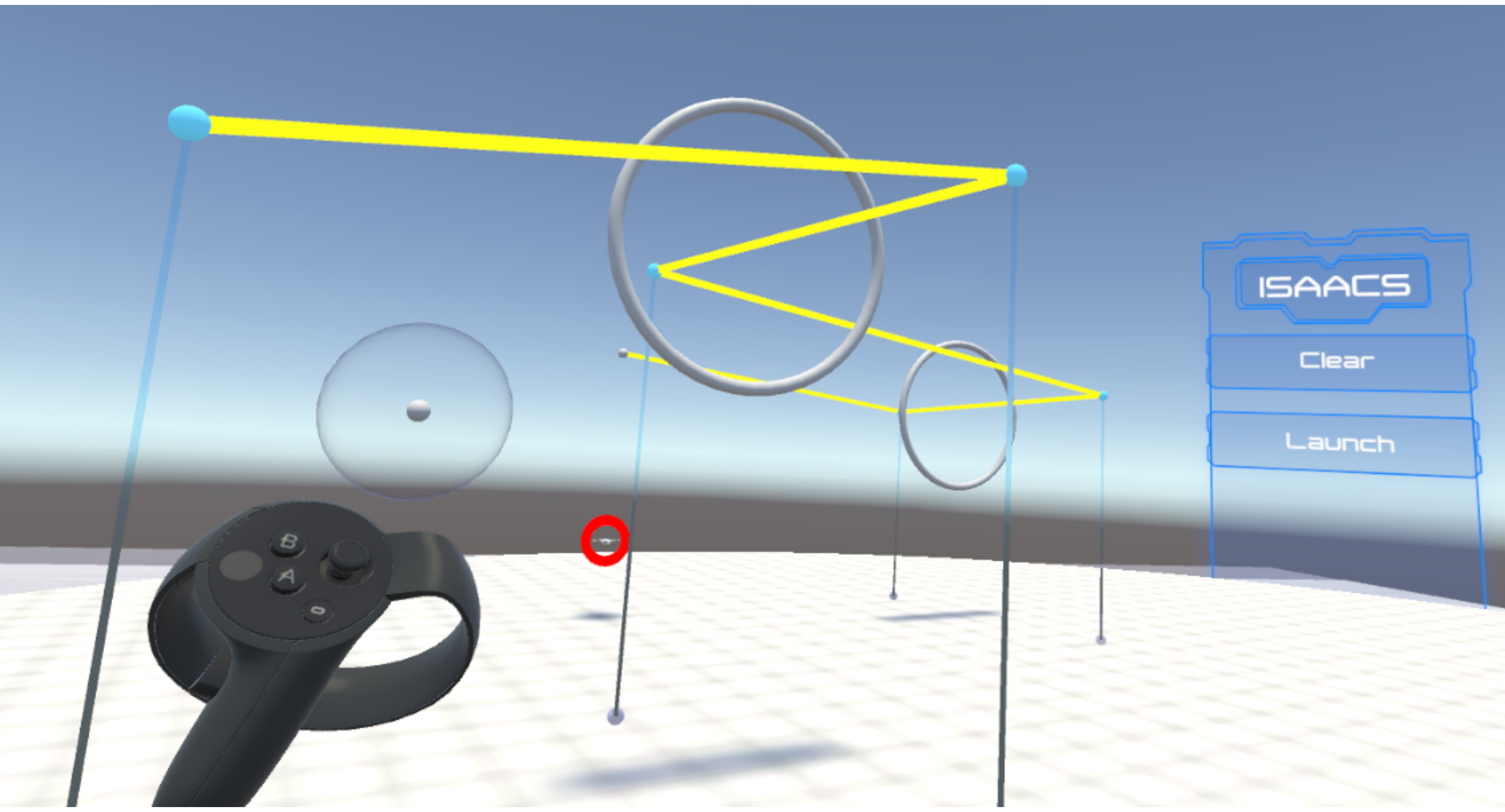}
    \caption{An example view from inside our VR interface. The user-defined path is displayed in yellow with a UAV beginning takeoff circled in red.} \label{fig:interface}
\end{figure}

In response to the usability challenge of manual controls, several commercial UAV systems allow non-expert users to bypass manual control and instead create flight paths using a tablet. These systems use automation to translate user-defined paths into low-level controls (i.e. roll, pitch, yaw, and thrust). Yet the 2D interface and the 2D input scheme impose usability challenges when the task requires precise 3D awareness and maneuvering. This work investigates the utility of an interface (shown in Figure \ref{fig:interface}) that combines immersive 6 DoF VR displays and 6 DoF inputs for controlling UAVs in such tasks.

\section{Related Work}

Human factors research shows that immersive displays improve performance on depth-critical tasks. The added depth cue provided by stereoscopic displays improves depth perception over traditional depth cues (e.g. perspective, shading, parallax, occlusion, and foreshortening) \cite{Lampton1995}. This improved depth perception provides particular benefit for height-sensitive tasks \cite{Scribner1998, Witmer1998} and precision-critical tasks \cite{Drascic}. In such tasks, stereoscopic displays reduce the gulf of evaluation as users are given more information to determine the state of the world and how close their goal is to the present state. However, McIntire et al. \cite{McIntire2014} suggest that the benefits of using immersive displays only hold for tasks that truly require depth. As such, it is unclear if the benefits of an immersive display will hold for the task of UAV path planning.

Immersive displays have been coupled with robotic operations to great effect. NASA rovers have been configured to reconstruct 3D models of their surroundings, which are viewed through head-mounted VR displays \cite{Nguyen2001}. This provides users with a better understanding of the remote environment where they are working and enables visualization of the vehicle's mechanics as it is commanded. A similar approach has been used to give construction teleoperators better situational and spatial awareness of a 3D task, albeit with a projected surrounding display instead of a head-mounted display\cite{Yamada2008}. Each application's spatial problem benefited from improved depth perception and a visualization of the environment.

The opportunities afforded by 3D interfaces have not been overlooked by the UAV research community. In 2006, Knutzon \cite{Knutzon2006} created the Battlespace Research Platform, a ground control system capable of real-time path re-planning and 3D immersive visualization through stereoscopic binoculars. However, interactions through a 2D game-pad interface limited users to selecting from a set of predefined paths or known locations visualized through the binoculars. Knutzon concluded his work by suggesting further investigation into whether a 3D immersive interface provided either a situational or general advantage over traditional 2D interfaces. Francesca et al. \cite{Francesca2009} developed an interface which displays the 3D virtual environment in which a UAV is deployed and allows destinations for the UAV to be designated via a 2D touch screen. They found that the 3D virtual display provided a clear and efficient understanding of the UAVs current state and environment. Honig et al. \cite{hoenig2015mixed} developed a system similar to ours, linking Crazyflie 2.0 quadrotor UAVs to a virtual environment using a motion capture system. This is used as a tool for evaluation of new algorithms, rather than as an interface for user drone control.

These prior applications have demonstrated the promise of immersive displays for spatial robotics problems. However, they have not combined these interfaces with visualized 6 DoF interactions. These interactions were employed by Jankowski et al. \cite{Jankowski2015}, who used a 6 DoF data glove alongside a first-person head-mounted VR display for the control of an inspection robot. This minimized difficulties in control and expanded the spatial perception  of the operator. These usability improvements in turn resulted in increased efficiency in the completion of 3D tasks over mono- and stereo-display interfaces. Industrial robot arms too have benefited from 6 DoF controllers' ability to set spatial paths \cite{chong2009robot,fang2012interactive,pettersen2003augmented}. Anderson et al. \cite{Anderson2015} created a 6 DoF remote for direct control of UAVs with directional haptic feedback to indicate movement towards high hazard areas.

Leveraging 6 DoF input capabilities reduces the gulf of execution as interactions can be contextualized within remote environments by directly moving input devices to corresponding relative positions in the virtual environment. This motivates an expectation of efficiency improvement when using 6 DoF interactions over 2D touch gestures for the task of 3D path planning for UAVs.

\subsection{Contribution}
We aim to validate that a VR interface designed for aerial path planning tasks can provide improvements in safety, efficiency and usability over traditional manual joystick control and 2D interfaces. We specifically examine three hypotheses:
 \begin{enumerate}
     \item \emph{Efficiency}: the 3D VR interface reduces path planning interaction time compared to a 2D touchscreen interface.
     \item \emph{Safety}: the 3D VR interface reduces crash rate for novice users compared to manual control or a 2D touchscreen interface.
     \item \emph{Usability}: A novice user will find the 3D VR interface easier to use compared to manual control or a 2D touchscreen interface.
 \end{enumerate}

Of the three hypotheses, Efficiency and Safety are objectively measured, while Usability is measured via a subjective survey. We compare these three measures for reproductions of commonly used UAV path planning schemes. Therefore, the functionality of the manual joystick control, 2D interface, and 3D VR interface are purposefully designed to be comparable.

In this paper we present an open-source aerial path planning platform in VR. The system uses 6 DoF head-mounted display (HMD) and 6-DoF inputs (via commercial products such as Oculus Rift and HTC Vive) to create a high-level waypoint-defined flight path for UAVs. This path is interpreted by a semi-autonomous control layer to provide low-level control inputs to the UAV, integrating sensor readings to accommodate external conditions. The open-source project will be linked upon publication.

Finally, we conduct a user study comparing our VR interface to a version of manual joystick control and a 2D tablet-based interface. This user study consists of 12 subjects, from which we draw a total of 6 hours of collected interaction data and 12 corresponding surveys. This evaluation exposes the potential benefits and drawbacks of using an immersive VR interface as a method of spatial communication between humans and robots.

\section{System Design and Implementation}
In this section, we discuss the UAV flight system used to interpret high-level commands from the test interfaces. We will also discuss the design of our VR interface, a comparable 2D tablet-based interface, and the hover-assisted manual control interface.

\subsection{UAV Flight System}
The UAV flight system is designed to potentially support multiple UAV and sensor types. In this paper, we choose to use a single Crazyflie 2.0 quadrotor UAV alongside OptiTrack visual telemetry for localization.\footnote{In other compatible configurations for outdoors, our system could work with larger UAVs (such as DJI Matrice 600) alongside application-ready telemetry sensing through RTK (Real-Time Kinematic) GPS.}

For the planning control module, we choose to
employ the Fast and Safe Tracking (FaSTrack) algorithm described in the work of Herbert et al \cite{herbert2017fastrack}.
This algorithm provides a principled decoupling between global path planning and local tracking control that provides bounds on how far a UAV may deviate from its planned trajectory. It should be noted that the FaSTrack algorithm is not critical to our implementation, and alternative UAV path following algorithms could be used with similar results.

Paired with the OptiTrack system for state updates in a laboratory setting, the FaSTrack system is our primary method for flying UAVs along given paths. A Robotic Operating System (ROS) API connects our interfaces to this control system to update the locations of the user's planned path points as well as launch/land commands. This ROS abstraction approach is similar to the one taken by Cheema et al. \cite{Cheema2016} in their 2D Control and Vision interface. This platform is used for all three test interfaces described in this paper.

\subsection{Virtual Reality Interface}

\begin{figure}[t]
    \centering
    \includegraphics[width=0.8\columnwidth]{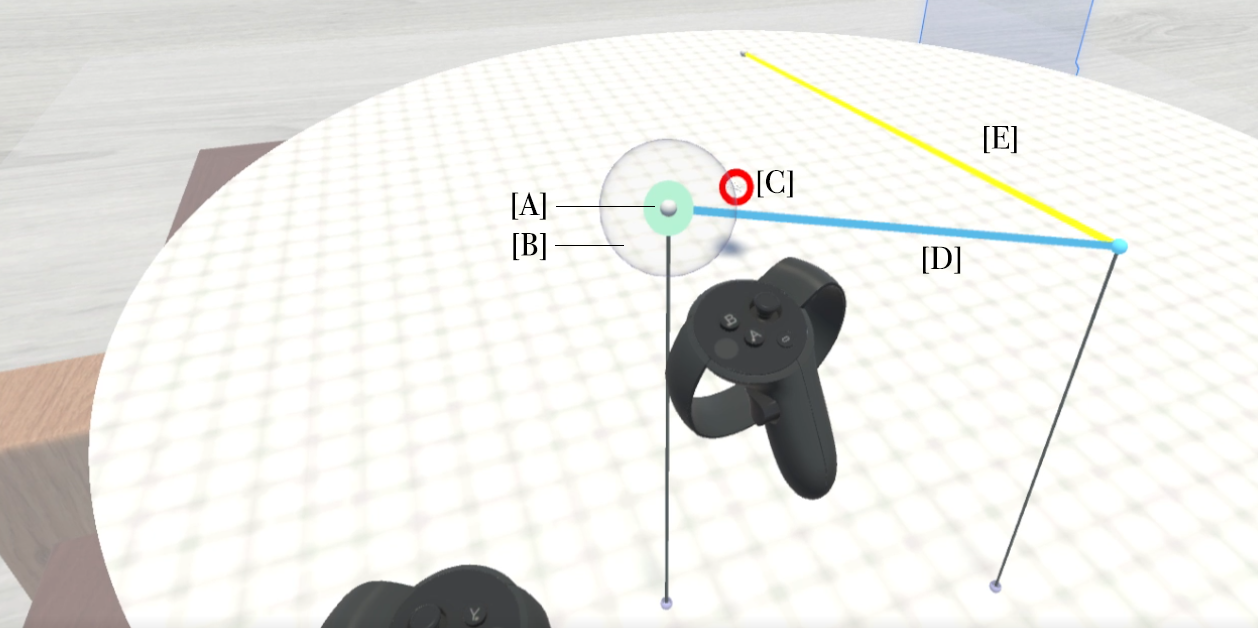}
    \caption{An example of the Direct Placement Method. The user uses the placement point [A] at the center of the selection zone sphere [B] in front of the controller as a guide to place waypoints. The drone [C] can be seen hovering in the background along with sections of the selected [E] and deselected path [D].}\label{fig:direct-placement}
\end{figure}

\begin{figure}[t]
    \centering
    \includegraphics[width=0.8\columnwidth]{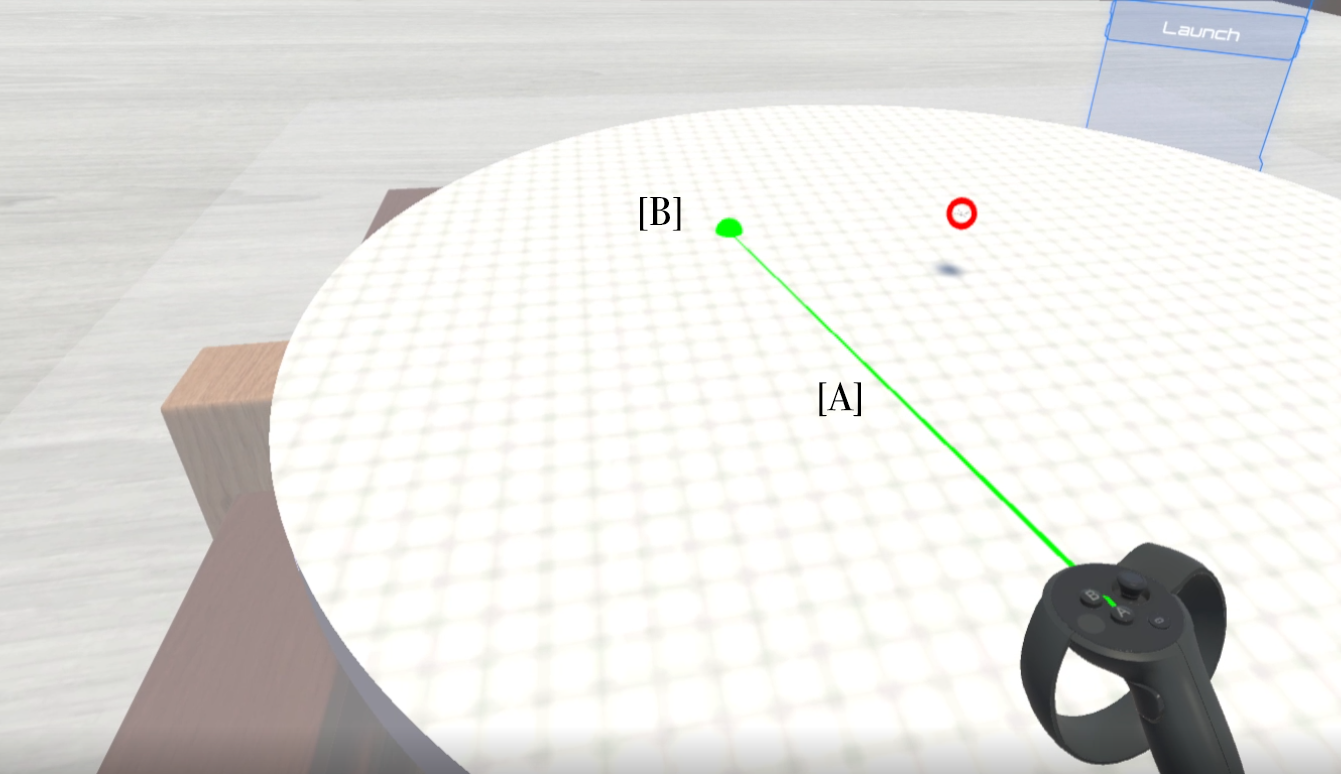}
    \caption{An example of the Secondary Placement Method. Using a green ray cast [A] from the controller's location, the user specifies an x-y location [B] to place a waypoint. Height is adjusted afterwards by tilting the controller upwards.} \label{fig:secondary-placement}
\end{figure}

We have developed the immersive VR interface in C\# using Unity 2017.2. We utilize the Oculus Rift CV1 headset and accompanying Touch controllers for 6 DoF display and inputs.

The interface is designed to display a 3rd person perspective of the virtual reconstruction of the environment that the UAV is deployed in. By giving operators a birds-eye view of the environment, the system allows them to monitor a UAV's current position and planning information without incurring the motion sickness often caused by first-person views with unpredictable motion, as seen in many other VR interfaces for UAVs \cite{Patrao2015, Luks2006}. In addition to adjusting the display view through direct head movement, the user is able to adjust their view of the environment using the touch controllers. By holding both triggers and pulling the controllers apart, the user can change the scale of the virtual environment. The user can also rotate the virtual environment about the vertical axis and translate the virtual environment along the horizontal plane using the joysticks. This facilitates viewing of and interaction with UAV flight trajectories from any position within the predefined operation area.

Within this camera view, a flight path is overlaid on top of the environment, consisting of waypoints and path lines connecting the waypoints. Sight lines extend below each waypoint to aid in pinpointing the exact location in the environment it is situated over. The colors for path lines and waypoints are chosen to stand out against the majority of environment textures.

These flight paths can be modified in two ways through the VR interface: a direct placement method and an indirect placement method. The direct placement method allows an operator to place waypoints at a fixed location relative to their right controller, which we denote as the placement point. The placement point is surrounded by a semi-transparent sphere (known as the selection zone) which allows the user to grab and adjust the placement of points within the zone, delete those points, or add new intermediate points to paths. Waypoints and path lines within the selection zone are highlighted to make it clear which elements will be modified by an action. Figure \ref{fig:direct-placement} illustrates an example of this direct placement method.

In addition to the direct placement method, there is an indirect placement method. This method allows the user to place waypoints at a distance by pointing the controller at a location at the base of the environment. A ray is cast outward from the controller to visualize where the controller is pointing. After selecting a location, the user tilts the controller upward to designate the height of the waypoint. This secondary placement method allows the user to quickly make changes to the path without needing to rescale or translate the environment model to put the placement area within arm's reach. Figure \ref{fig:secondary-placement} illustrates an example of this secondary placement method.

\subsection{2D Interface} \label{sec:2D}
\begin{figure}[h]
    \centering
    \includegraphics[width=0.8\columnwidth]{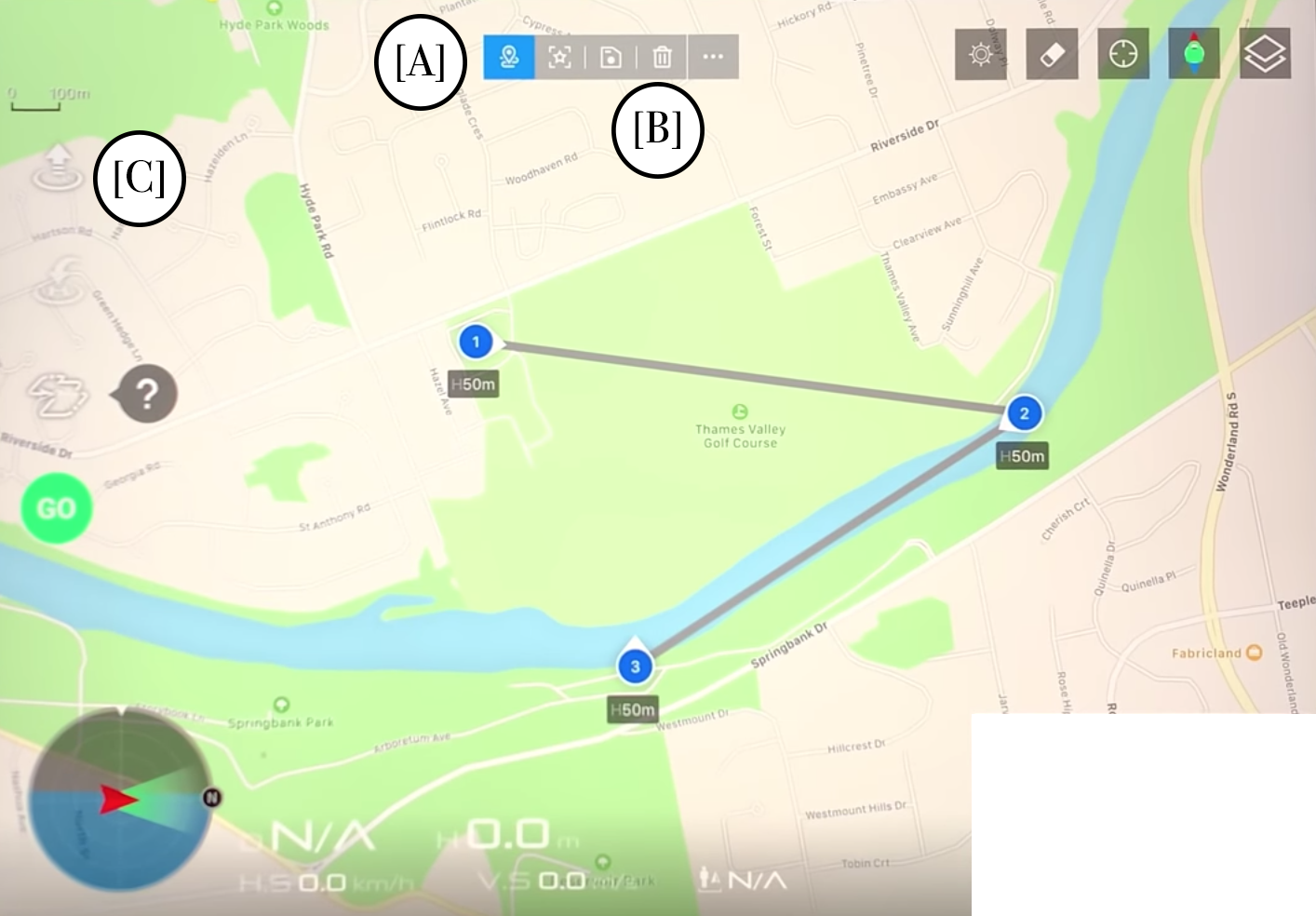}
    \caption{Illustration of DJI GO Interface. Waypoints are placed by tapping [A], the waypoint placement button, then tapping on the screen at the desired location. Swiping up or down allows the user to set the height. Deleting a waypoint is done by tapping [B] the trash can icon and tapping the waypoint to be destroyed. Take off and landing is done through [C] the top button on the left side of the screen. Website: https://www.dji.com/goapp }
    \label{fig:DJI-GO}
\end{figure}

To fairly validate our hypotheses against typical 2D UAV interfaces, we have created a Crazyflie-compatible 2D interface to reflect current industry benchmarks such as the DJI GO app (as shown Figure \ref{fig:DJI-GO}). The 2D interface is developed in C\# using Unity 2017.2, incorporating the open-sourced TouchScript software for touchscreen interactions and hosted on a Microsoft Surface Pro.

\begin{figure}[h]
    \centering
    \includegraphics[width=0.8\columnwidth]{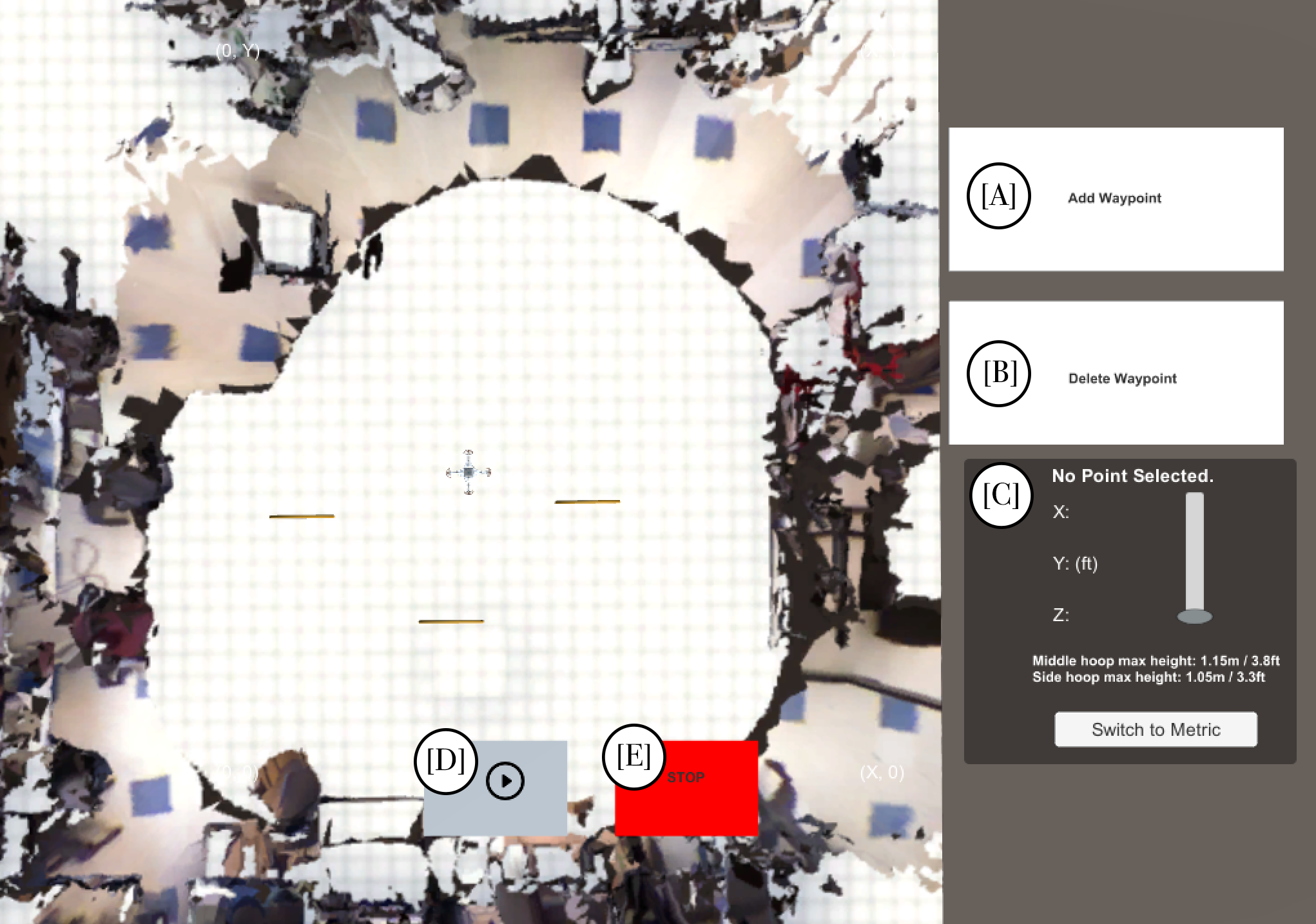}
    \caption{Illustration of our 2D Interface at startup. Waypoints are placed by tapping [A], the 'Add Waypoint' button, then tapping on the screen at the desired location and setting the height using [C], the right-hand side slider. Deleting a waypoint is done by tapping [B], the delete waypoint button, and then tapping the waypoint to be destroyed. Take off and landing is done through [D], the play button and stop button on the bottom of the screen. The UAV and hoops in the interface are tracked in real-time. Buttons and the slider allow users to place and delete waypoints, adjust waypoint height, and launch or land the UAV.}
    \label{fig:touch-interface}
\end{figure}

Our interface (Figure \ref{fig:touch-interface}) incorporates all features from DJI GO that are necessary for the execution of tasks that we aim to evaluate in this study. Specifically, we include waypoint placement, height setting, location adjustments, and deletion of waypoints under a bird's-eye view. In replicating these features, we postulate that we can fairly compare user performance on our interface to that of interfaces available on the market.

In the 2D interface, the operator is presented with a bird's eye view of the virtually reconstructed environment, which is denoted as the map. The height measurements of waypoints and objects are also displayed in the map as a substitute for the depth perception of the immersive interface. One or two finger touchscreen gestures can be used for panning, scaling, and rotating, as utilized by the DJI Go interface.

For the sake of parity, the UAV hardware and the planning control module used by the 2D interface are identical to those used for the VR interface. To place waypoints, the user is prompted to select a 2D location on the map followed by a height input, which is determined by a slider (Figure \ref{fig:touch-interface}). Deletion and modification are achieved, respectively, by toggling the Delete Waypoint button then selecting a waypoint and by dragging any existing waypoint.

\begin{figure}[h]
    \centering
    \includegraphics[width=0.9\columnwidth]{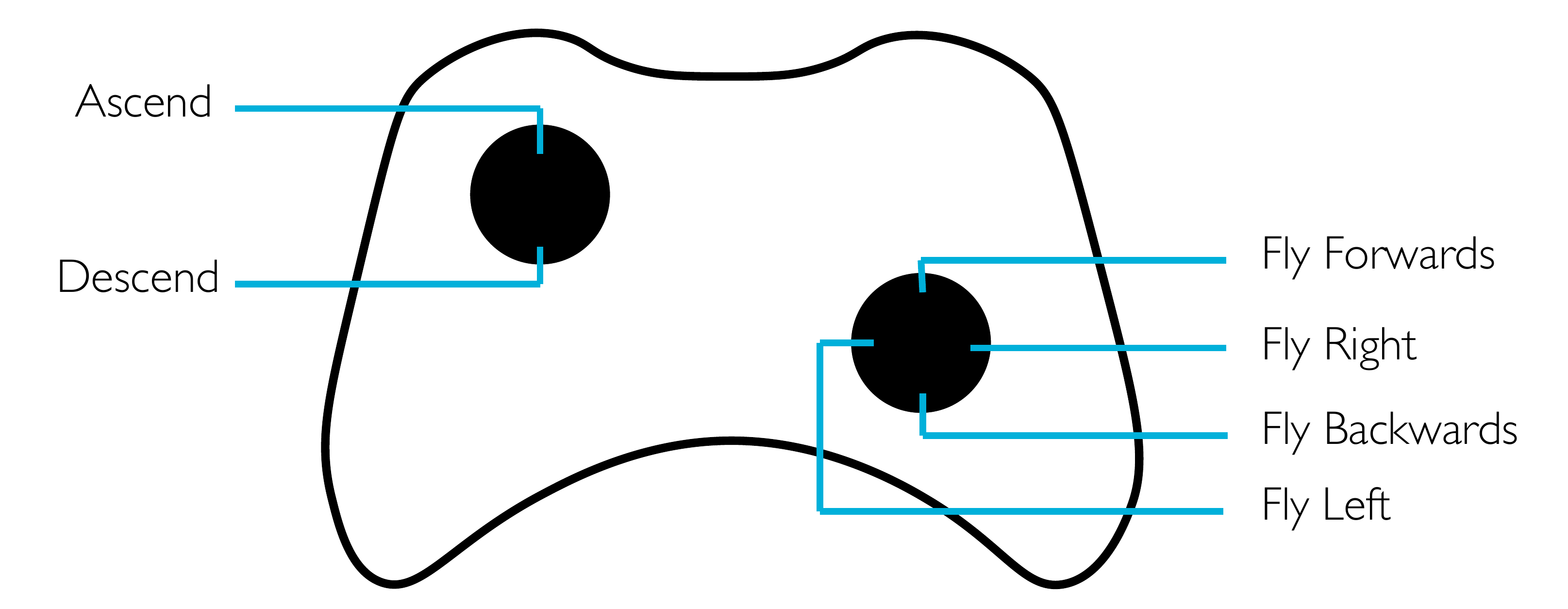}
    \caption{Illustration of the manual control interface we use for comparison. The user has direct control over the position of the UAV using the controller's two joysticks.} \label{fig:controller}
\end{figure}

\subsection{Manual Control}
Another standard UAV interface is for a pilot to command velocities in real-time with close supervision (typically within line-of-sight). One joystick is used to control the UAVs' horizontal planar movement and a second joystick controls the vertical thrust (as depicted in Figure \ref{fig:controller}).
For fair comparison, the manual controller also benefits from the Optitrack localization-based stabilization that is used in the FaSTrack-hierarchical systems. Rather than setting a waypoint, as in the 2D and VR interfaces, the tracking point is directly offset from the UAV's current position depending on the joysticks' positions. Varying linearly up to one meter at 100\% joystick deflection with a stabilizing dead-zone from 0 to 10\% deflection. Given no input, the UAV will autonomously hover at its current setpoint. This removes the burden from the user to actively stabilize the UAV at a given point.

\section{Evaluation}
Our experiment evaluates the efficiency, safety, and usability of the three interfaces by observing human users complete a series of hoop navigation tasks. These tasks are meant to be representative of applications requiring UAVs to navigate small spaces, accurately follow a path, and avoid obstacles. 

\subsection{Participants}

Participants for this study were pooled from university mailing lists related to Computer Science and VR. Of our 12 participants, three were female and nine were male. The age of our participants ranged from 19 to 31. Acknowledging that prior VR exposure poses a risk of confounding our results, we attempted to achieve a distribution of VR-related experience among the participant group. Out of 12 participants, 6 participants reported little or no prior experience, 3 reported extensive prior experience, and the final 3 reported having moderate VR-related experience. 

\subsection{Experimental Design}
We perform a within-subjects experimental design across the three interface types, counterbalancing the order of exposure to avoid familiarity-based hysteresis confounds. We test a mixture of both novice and experienced VR users with six possible randomized orderings of exposure to the three interfaces. The user study is administered by two interviewers, with one reading the experiment script and the other preparing the system and providing materials.

\begin{figure}[h!]
    \centering
    \includegraphics[width=0.9\columnwidth]{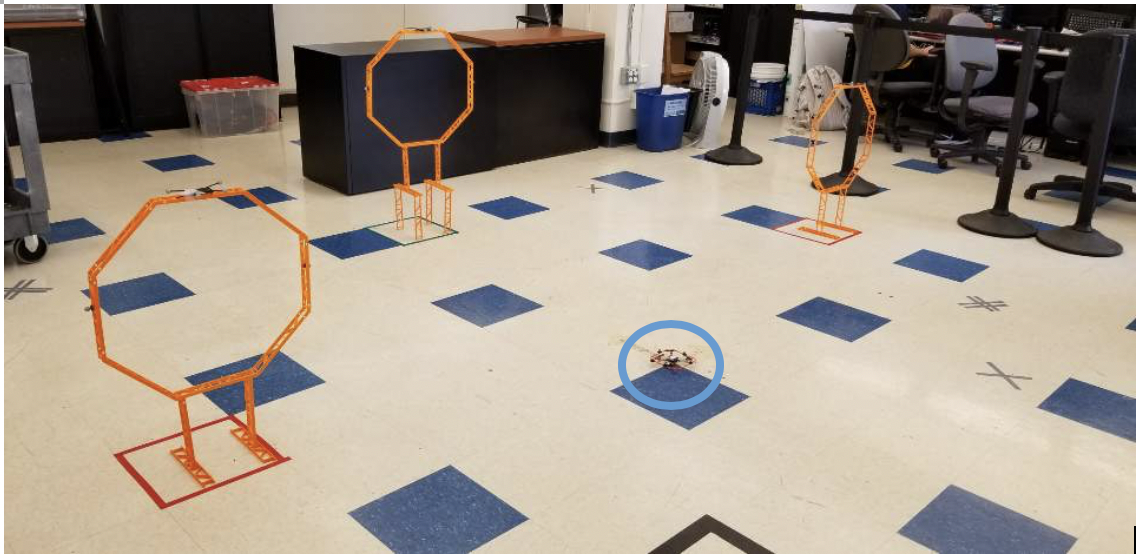}
    \caption{Lab configuration for an experiment with three hoops set up at different heights as obstacles. The miniature Crazyflie UAV (circled in blue) can fit through these hoops with approximately a body length on either side.} \label{fig:lab-photo}
\end{figure}

The subjects are tasked with tele-operating a single UAV through a spatial navigation task. Three circular hoops of different heights, marked as ``A'', ``B'', and ``C'', are positioned throughout the 10ft by 10ft by 10ft laboratory space (Figure \ref{fig:lab-photo}). The task is to fly the UAV through a given sequence of hoops without crashing. For this task, "crashing" is defined in two ways. The first is collision of a UAV rotor with the hoop. The second is collision of any part of the UAV with the hoop, followed immediately by collision of any part of the UAV with the ground. These definitions were given to each participant before each set of trials. The UAV battery is replaced before every new set of tasks . The obstacles' locations are marked by a square area to which the hoop stands are aligned for consistency. 

The subjects are introduced to each interface via a verbal tutorial script and short demo. In the case of the 2D interface, the demo is live and carried out by one experimenter. The experimenter introduces one interface feature at a time, asking the participant to use each feature to show adequate understanding (allowing for questions if necessary). In the case of the virtual reality interface, the demo is automated in Unity. This demo parallels the live 2D demo and uses verbal instruction playback to mimic the experimenter and an automated script to confirm understanding of the controller functions. For the manual control interface, a verbal explanation of the button configurations is given along with a sheet illustrating the control layout. During the explanation the user is able to observe the movement of the UAV as he or she operates the controller in real time. Immediately following each interface's respective tutorial, the subjects are permitted to practice during a 3 minute familiarization phase. After each tutorial and practice phase is completed, the subjects perform three experiment runs for that interface. Each run requires a different sequence of hoop traversals and last until the user either fails (by crashing the UAV) or succeeds (by completing the traversals). 

\begin{figure}[h]
    \centering
    \includegraphics[width=0.9\columnwidth]{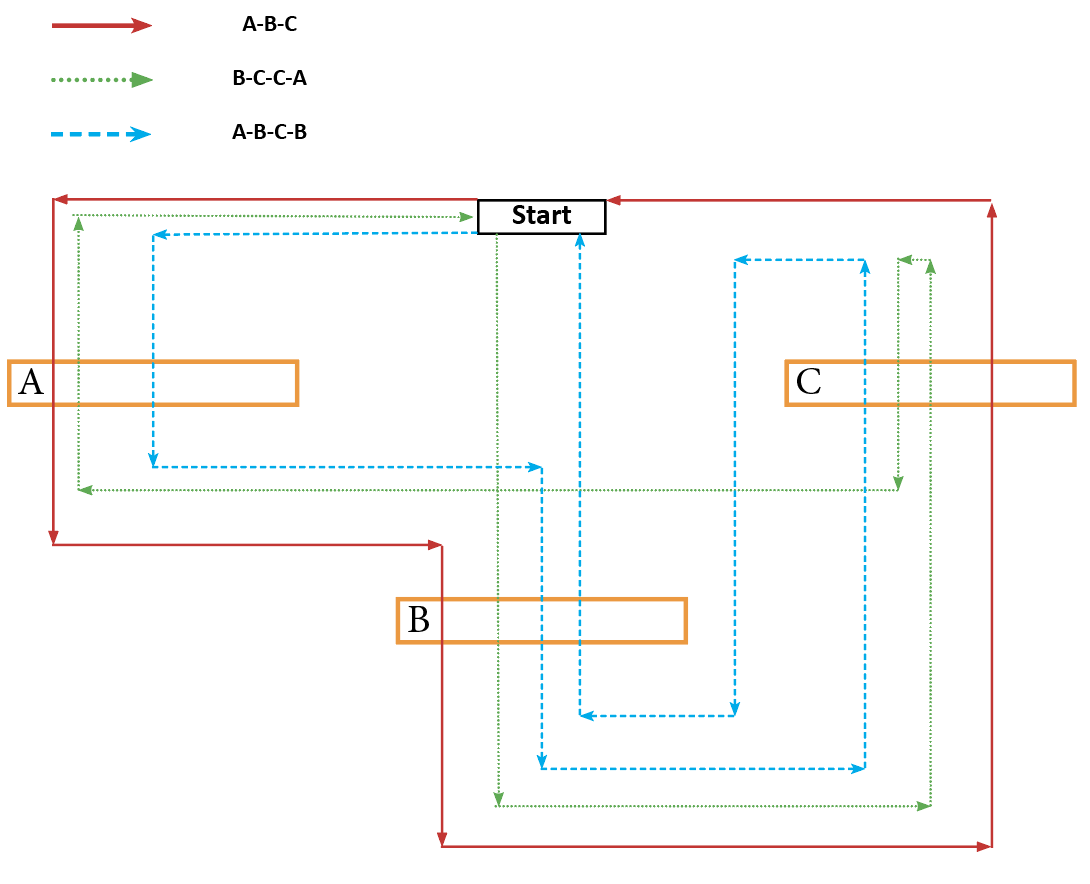}
    \caption{Example of a valid path for each trial, with each path denoted by a different color.} \label{fig:ABC-path}
\end{figure}

The path order of the hoops for each set of trials are shown in Figure \ref{fig:ABC-path}. The path A-B-C-B denotes any path that goes through hoops A, B, C, then B again, in that order. The traversal of each hoop can be completed from either side. The paths and path orders remain constant among all interfaces. 

Following the end of the full experiment with all interfaces, a survey is presented to the subject, asking them to rate their agreement with a series of five statements for each interface.  Each statement is reported on a 7-point Likert scale. The first four statements are designed to measure usability while the fifth statement measures how pleasant the interface is to use. The statements are shown
in Table \ref{tab:surveyQuestions}. Note that the fourth question reverses the valence to test if users are paying attention. In the analysis, this reverse question's scores are inverted to match the other scales. 

\begin{table}[tb]
\vspace{10pt}
\renewcommand{\arraystretch}{1.1}
\caption{Survey Questions}
\label{tab:surveyQuestions}
\centering
\begin{tabular}{|l | c|}
\hline
\hspace{18mm}Question & Score \\
\hline
The robots and I worked fluently&\\ together using this interface. & 1-7 \\
\hline
I quickly adapted to this interface. & 1-7 \\
\hline
Controlling the system with&\\ this interface came naturally to me. & 1-7 \\
\hline
This interface was confusing. & 1-7 \\
\hline
This interface was pleasant to use. & 1-7 \\
\hline
\end{tabular}
\vspace{-10pt}
\end{table}

\section{Results and Discussion}
\begin{figure}[h!]
    \centering
    \includegraphics[width=0.9\columnwidth]{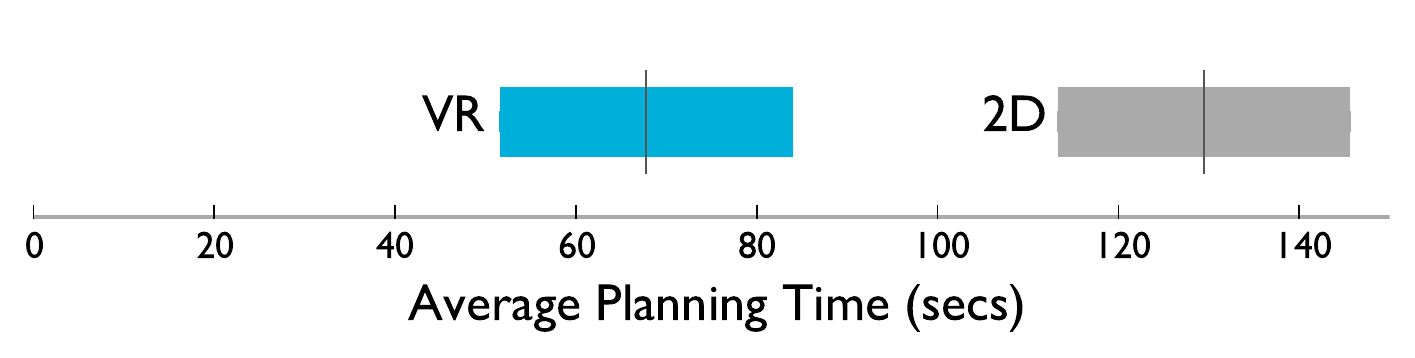}
    \caption{The mean planning times for 2D and VR spatial interfaces with sharply separated 95\% confidence intervals (shaded).}
    \label{fig:planningtime}
\end{figure}

To test the efficiency of the interface schemes, a one-way repeated-measures Analysis of Variance (ANOVA) compares the planning time of the 3D immersive interface to the 2D touch-based baseline. As shown in Figure \ref{fig:planningtime}, the VR interface demonstrates a decrease of mean planning time from 129 seconds to 68 seconds with strong statistical significance ($F(1,11) = 60.2542, p < 0.0001$). There are two factors that may contribute to this reduction in planning time. One is that the increased depth perception and viewpoint adjustment afforded by the VR interface allow the user to more quickly determine a viable path for the drone to fly. The other factor is that the 6DoF controller allows the user to specify the determined path in reduced time. Due to the simplicity of the planning task in this experiment, we believe the latter factor to be the larger contributor to this reduction. 

\begin{figure}[h!]
    \centering
    \includegraphics[width=0.9\columnwidth]{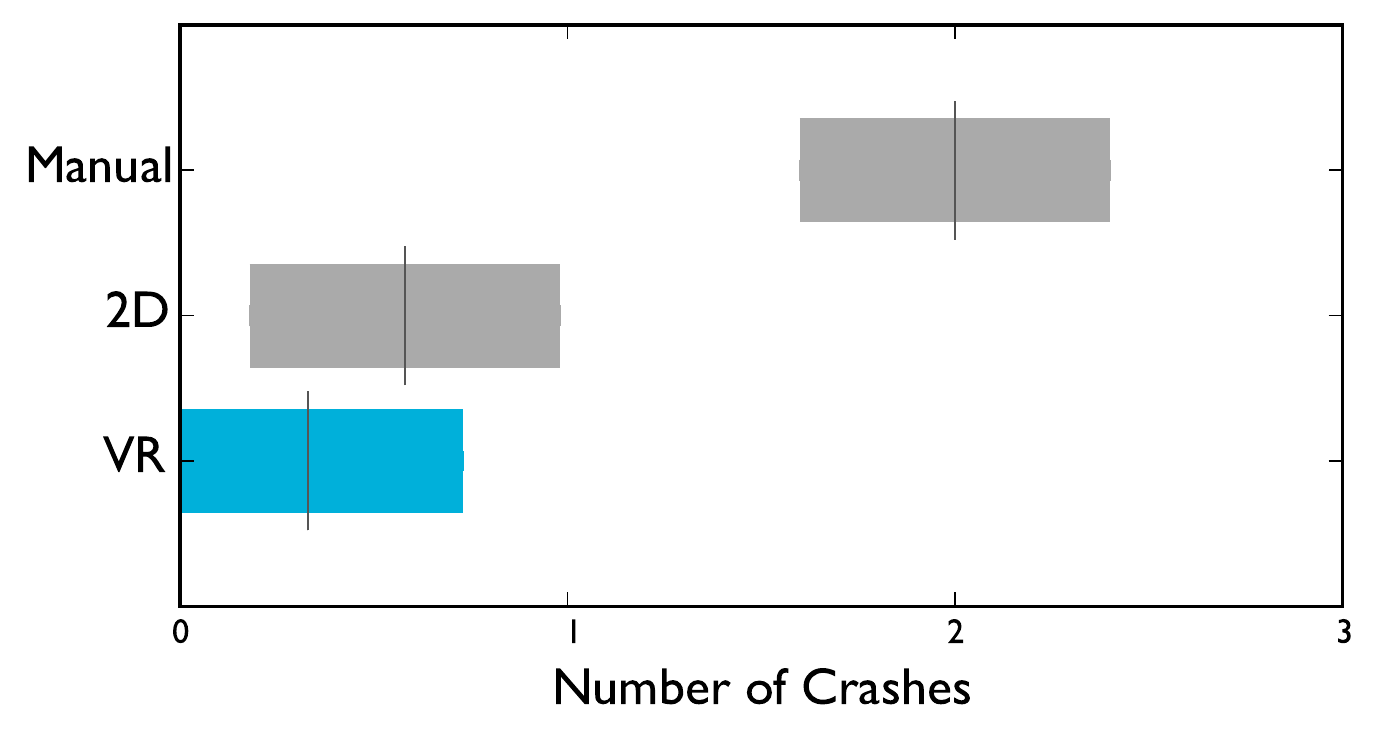}
    \caption{Mean number of crashes plotted with 95\% confidence intervals shaded in. The overlapping spatial interfaces have indiscernible improvements, while the manual controller is statiscally significantly ($p<0.0001$) less safe with almost twice the number of crashes}
    \label{fig:crashes}
\end{figure}

A similar one-way repeated-measures ANOVA tests Hypothesis 2: the number of collisions will drop significantly for the immersive interface as compared to the 2D touchscreen interface or the manual controller. The user study data (shown in Figure \ref{fig:crashes}) contains a significant effect on the number of collisions ($F(2,22) = 17.86, p < 0.0001$). An all-pairs post-hoc Tukey analysis shows that the virtual reality interface and 2D touch interface statistically significantly reduce crashes ($p < 0.0003$) over the manual interface. No statistical difference is demonstrated between the two spatial interfaces ($p = 0.6881$). This suggests that the increase in safety is due to the hierarchical spatial planning that decouples safety control from user path-planning. This may also indicate that this task as designed is not one that truly requires perception of depth in the display itself and users were able to accurately judge the proper height for the drone to fly through observation of the scene. This may change in the scenario where the user is not co-located in the scene, but that was not tested by this experiment. 

\begin{figure}[h!]
    \centering
    \includegraphics[width=0.9\columnwidth]{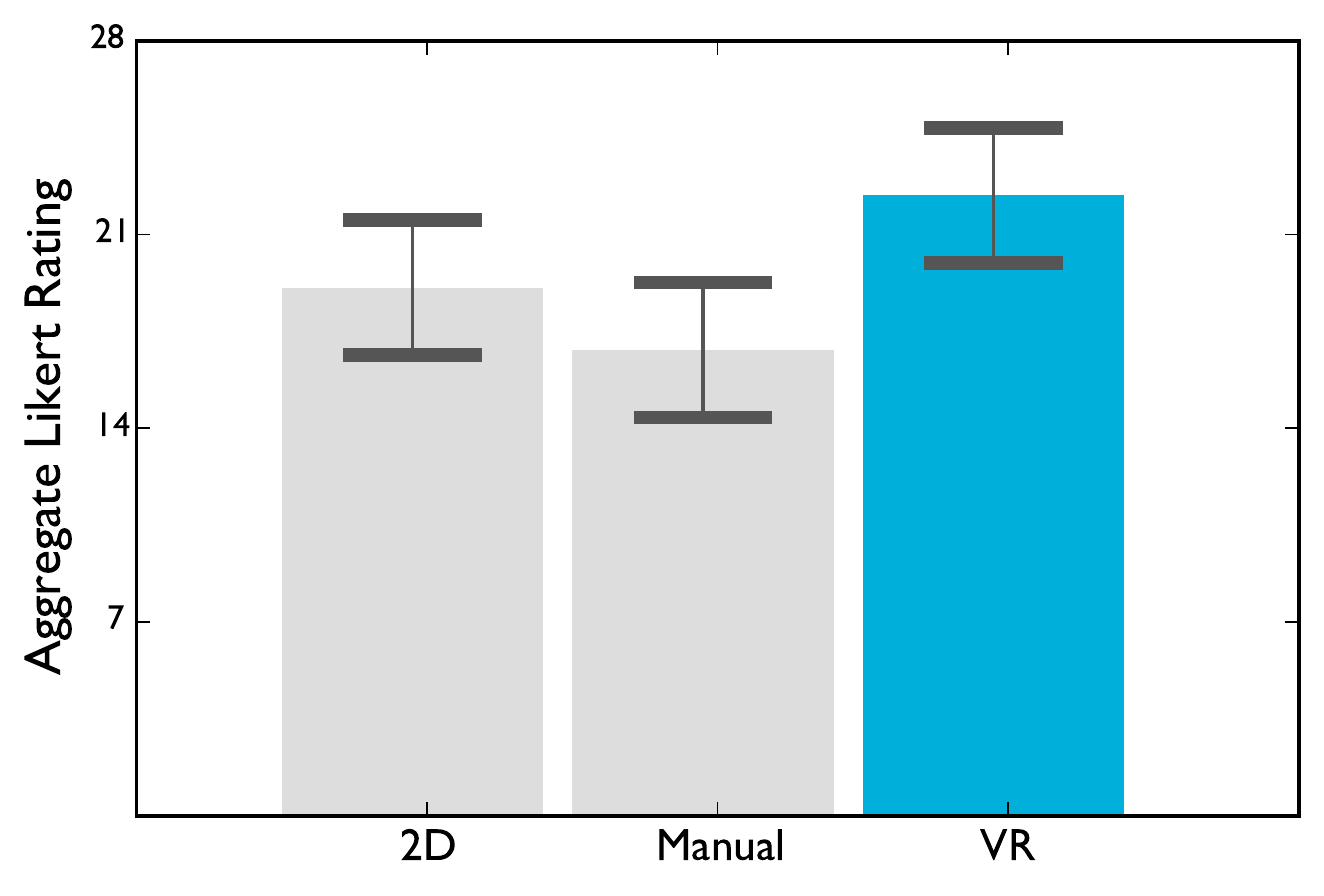}
    \caption{The three interfaces compared by average subjective scores. The scores aggregate subject's degree of agreement with four statements about fluency, confusion, naturalness, and ease of adaptation which measure the same factor with Cronbach alpha = 0.7. The plotted confidence intervals show how the VR interface improved over manual control with statistical significance.\label{fig:SurveyResults}}
\end{figure}

The post-experiment comparison survey measures the interface usability. To test the construct validity of this subjective measure, we apply the Cronbach's alpha amongst the usability survey questions to calculate an internal consistency of 0.7047. Therefore the survey questions can be acceptably interpreted as measuring the same parameter and are aggregated into a single measure of usability to mitigate measurement noise. A one-way repeated-measures ANOVA on this joint measure shows that the VR interface improves an average 5.58 Likert points over the manual controller (with statistical significance $p = 0.0159 < 0.05$), while all other pairings are statistically indiscernible as shown in Figure \ref{fig:SurveyResults}. While this short survey certainly does not cover all aspects of usability, it does demonstrate a preference for the VR interface over Manual control. It also indicates that, despite the potential unfamiliarity of VR, the immersive interface does not introduce excessive confusion for these users as compared to the 2D interface for this task. Across all three metrics, we did not observe a statistically significant difference in the performance of the VR interface when comparing those users with VR experience to those users without VR experience. It should be noted however that, given the limited age range of our test subjects, these results may not hold outside this population.

\section{Conclusions and Future Work}
In this paper, we have demonstrated that a VR interface can provide a safe and more efficient alternative to 2D touchscreen interfaces for 3D UAV path planning tasks. Our interface provides comparable safety and usability improvements over manual interfaces while significantly reducing path planning interaction time compared to a 2D touchscreen interface. 

In the future, to further this research, we plan to evaluate how the benefits of this interface can extend to multiple UAV fleet monitoring and how we can allow the underlying controller to provide more information to the user.

As UAV capabilities begin to support more complex tasks in 3D space, user interfaces will also have to adapt to support these challenging new tasks. 
Virtual reality interfaces present a promising solution for the challenges introduced by these new UAV control tasks.

\balance{}

\balance{}

\bibliographystyle{SIGCHI-Reference-Format}
\bibliography{sample}

\end{document}